\documentclass[aps,prx,a4paper,amssymb,amsmath,reprint,superscriptaddress]{revtex4-1}
\usepackage{amsmath,amssymb}
\usepackage{bm,braket,latexsym,multirow}
\usepackage{bbding,pifont,wasysym}
\usepackage{colortbl,array}
\usepackage[dvipsnames]{xcolor}

\usepackage{graphicx}

\usepackage[
pdftex,
draft=false,
bookmarks=true,
bookmarksnumbered=false,
bookmarksopen=false,
bookmarksopenlevel=4,
colorlinks=true,
anchorcolor=black,
citecolor=blue,
filecolor=black,
linkcolor=black,
linkbordercolor={1 0 0},
urlcolor=violet,
pdfborder={0 0 1},
pdftitle={},
pdfauthor={},
pdfsubject={},
pdfkeywords={},
pdfpagemode=UseThumbs,
]{hyperref}

\usepackage{xspace}
\newcommand{\Pa}{$\mathcal{P}$\xspace}
\newcommand{\T}{$\mathcal{T}$\xspace}
\newcommand{\PT}{$\mathcal{PT}$\xspace}

\newcommand{\mr}[2]{#1 \, {\mathrm{ \,  #2 \,}}}
\newcommand{\bk}{{\bm{k}}}
\newcommand{\bA}{{\bm{A}}}

\newcommand{\limvec}{\lim_{\bm{A}\rightarrow \bm{0}}}
\newcommand{\utt}{UTe$_2$}
\newcommand{\vj}{J}

\begin{document}

\title{Nonreciprocal Meissner Response in Parity-Mixed Superconductors}
\author{Hikaru Watanabe}
\affiliation{RIKEN Center for Emergent Matter Science (CEMS), Wako 351-0198, Japan}
\affiliation{Department of Physics, Graduate School of Science, Kyoto University, Kyoto 606-8502, Japan}
\author{Akito Daido}
\affiliation{Department of Physics, Graduate School of Science, Kyoto University, Kyoto 606-8502, Japan}
\author{Youichi Yanase}
\affiliation{Department of Physics, Graduate School of Science, Kyoto University, Kyoto 606-8502, Japan}
\affiliation{Institute for Molecular Science, Okazaki 444-8585, Japan}
\date{\today}

\begin{abstract}
The parity breaking gives rise to rich superconducting properties through the admixture of even and odd-parity Cooper pairs. A new light has been shed on parity-breaking superconductors by recent observations of nonreciprocal responses such as the nonlinear optical responses and the superconducting diode effect. In this Letter, we demonstrate that the nonreciprocal responses are characterized by a unidirectional correction to the superfluid density, which we call nonreciprocal superfluid density. This correction leads to the nonreciprocal Meissner effect, namely, asymmetric screening of magnetic fields due to the nonreciprocal magnetic penetration depth. Performing a microscopic analysis of an exotic superconductor \utt{} and examining the temperature dependence and renormalization effect, we show that the nonreciprocal Meissner effect is useful to probe parity-mixing properties and gap structures in superconductors.

\end{abstract}
\maketitle

\textit{Introduction---}
The nonlinear and nonreciprocal responses are recently attracting interest in various fields of condensed matter physics. For instance, second harmonic generation and photocurrent creation have been applied to a probe of the symmetry breaking in matters and the topological nature of electrons~\cite{Fiebig2005-ne,Orenstein2021,Ideue2021}.
Recent studies explored the nonreciprocal responses in superconductors. The nonreciprocal electric conductivity~\cite{Tokura2018}, a rectified conductivity originating from the parity violation, is strongly enhanced by the superconducting fluctuation~\cite{Wakatsuki2017} and by the vortex dynamics~\cite{Hoshino2018,Zhang2020}. Furthermore, the recent efforts have clarified the nonreciprocal superconducting phenomena such as the nonreciprocal critical current~\cite{Ando2020,yuan2021supercurrent,daido2021intrinsic,he2021phenomenological,Ilic2021} and Josephson current~\cite{baumgartner2021josephson,wu2021realization,he2021phenomenological}.
The nonreciprocal critical current realizes the superconducting diode effect, indicating that the electrical resistivity is zero in a direction while finite in the opposite direction.
The nonreciprocal optical responses have also been observed in superconducting systems whose parity violation stems from the spontaneous order intertwined with superconductivity~\cite{Zhao2017-ka,Lim2020,De_la_Torre2020} or the injected supercurrent~\cite{Yang2019,Nakamura2020,Vaswani2020}. Building on the superconducting properties, the nonreciprocal phenomena imply richer functionalities.

There is a symmetry requirement of nonreciprocal responses which is unique to superconductors.
In addition to the space-inversion symmetry breaking, its combination with the gauge symmetry has to be broken. 
In particular, the odd-parity superconductivity is insufficient to cause the nonreciprocal response because the above symmetry holds. 
Superconductors must have no definite parity under the \Pa{} operation to host nonreciprocal responses, implying the \textit{parity-mixed superconducting state}.
Conversely, the nonreciprocal response may be an indicator of the parity mixing in superconductors as proposed in the prior study of the nonreciprocal conductivity~\cite{Wakatsuki2018}.
This potential indicator of the parity-mixed superconducting state may allow us to identify the relation between the parity violation and superconducting symmetries which has been intensively investigated with the noncentrosymmetric superconductors such as CePt$_3$Si~\cite{NCSC_book}.

Considering the high interest in the research community, it is desirable to further explore nonreciprocal properties of superconductors.
While previous theoretical studies have focused on DC or low-frequency charge transport~\cite{Wakatsuki2017,Hoshino2018,Wakatsuki2018,daido2021intrinsic,yuan2021supercurrent,he2021phenomenological}, nonreciprocal nature may appear in the other responses as well.
Recently the authors have identified anomalous contributions to the nonlinear optical conductivity in superconductors, which diverges in the low-frequency limit~\cite{WatanabeDaidoYanase} as is the case for the linear optical conductivity~\cite{Tinkham1959SCskin}.
This in turn implies an inherent directionality in the Meissner response, i.e., the \textit{nonreciprocal Meissner effect.}
Since the Meissner effect plays the central role in superconductivity, the nonreciprocal Meissner effect may also contribute to a deeper understanding of the parity-breaking superconducting states.

                \begin{figure}[htbp]
                \centering
                \includegraphics[width=0.95\linewidth,clip]{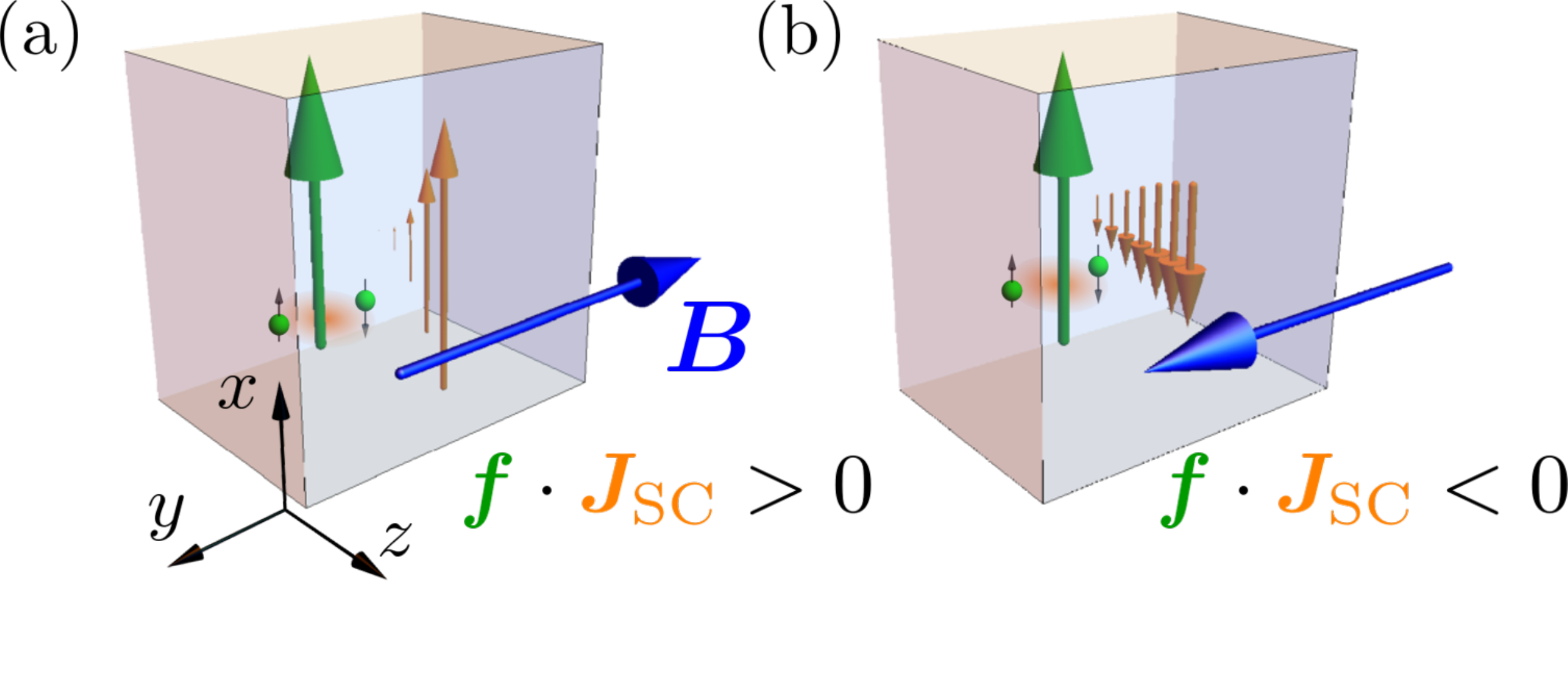}
                \caption{Nonreciprocal Meissner effect. The nonreciprocal superfluid density $\bm{f}$ (green arrow) (a) strengthens or (b) diminishes the  supercurrent $\bm{J}_\text{SC}$ (orange arrows) shielding the external magnetic field (blue arrow).}
                \label{Fig_schematic}
                \end{figure}

This Letter consists of two parts.
First, we show that the anomalous nonlinear conductivity arising from the parity-mixing leads to the unidirectional correction to the rigidity of the superconducting state, which we call \textit{nonreciprocal superfluid density}.
Various nonreciprocal responses of superconductors are characterized by the nonreciprocal superfluid density.
Second, to demonstrate exotic phenomena arising from the nonreciprocal superfluid density, we elaborate the nonreciprocal property of the Meissner effect (Fig.~\ref{Fig_schematic}).
With the microscopic analysis implementing the model of a putative parity-mixed superconductor \utt{}, we show that the nonreciprocal Meissner effect is sensitive to the parity violation and hence applicable to detection of the parity-mixed superconducting state.

\textit{Nonreciprocal property of the superfluid density ---}
The nonreciprocal electric conductivity of superconductors has recently been formulated by using the density matrix and Green function methods~\cite{WatanabeDaidoYanase}. The leading nonreciprocal correction to the electric current is given by the second-order component
		\begin{equation}
                J_\alpha^{(2)} (\omega) = \int \frac{d\Omega}{2\pi} \sigma_{\alpha;\beta\gamma} (\omega,\Omega) E_\beta (\Omega) E_\gamma (\omega- \Omega).
                \end{equation}
In the low-frequency limit, we obtain the nonlinear optical conductivity
                \begin{align}
                \sigma_{\alpha;\beta\gamma}(\omega,\Omega) 
                        &= \frac{f_{\alpha\beta\gamma}}{2\Omega (\omega-\Omega)} \notag \\
                        &-\frac{i}{4}  \limvec \left( \frac{1}{\omega-\Omega} \partial_{A_\beta} \sigma_{\alpha\gamma}^{(\bA)} + \frac{1}{\Omega} \partial_{A_\gamma} \sigma_{\alpha\beta}^{(\bA)}   \right). \label{anomalous_photocurrent_Green_function_method}
                \end{align}
The diverging terms proportional to $\Omega^{-2}$ or $\Omega^{-1}$ are unique to the superconducting state while they are forbidden in the normal state~\cite{Michishita2020}.
Thus, we call the terms anomalous contributions.
We suppressed $O(\Omega^0)$ terms comprising non-divergent nonlinear optical conductivity 
since it is negligible in the low-frequency regime.

The anomalous nonreciprocal conductivity in Eq.~\eqref{anomalous_photocurrent_Green_function_method} is determined by the nonreciprocal superfluid density $f_{\alpha\beta\gamma}$ (NRSF) and the conductivity derivative $\partial_{A_\gamma} \sigma_{\alpha\beta}^{(\bA)}$. The NRSF is given by
		\begin{equation}
                f_{\alpha\beta\gamma} = \limvec \partial_{A_\alpha}  \partial_{A_\beta} \partial_{A_\gamma} F_{A},
        \end{equation}
where $F_A$ is the free-energy obtained from the Hamiltonian containing the vector potential $\bA$. Since the superfluid density is given by $\rho^\text{s}_{\alpha\beta} = \limvec \partial_{A_\alpha}  \partial_{A_\beta} F_{A}$, the NRSF is regarded as a unidirectional correction to $\rho^\text{s}$. When the superfluid density is isotropic $\rho^\text{s}_{\alpha\beta} = \rho^\text{s} \delta_{\alpha\beta}$, the NRSF is recast as the vector $\bm{f}$ which has the same symmetry as the electric current and the toroidal moment~\cite{Spaldin2008,Hlinka2014}. Thus, the NRSF is allowed in the absence of both \Pa{} and time-reversal (\T{}) symmetries.

In the second term of Eq.~\eqref{anomalous_photocurrent_Green_function_method}, $\sigma_{\alpha\beta}^{(\bA)}$ denotes the regular part of the linear static conductivity calculated with the Bogoliubov-de Gennes Hamiltonian including $\bA$. Its derivative $\partial_{A_\gamma} \sigma_{\alpha\beta}^{(\bA)}$, which we call conductivity derivative, is decomposed into the symmetric and antisymmetric parts in terms of the permutation of indices $(\alpha, \beta)$ for $\sigma_{\alpha\beta}^{(\bA)}$. 
The symmetric part of $\sigma_{\alpha\beta}^{(\bA)}$ corresponds to the Drude contribution, while the anti-symmetric part is the Berry curvature term.
Thus, we call the corresponding components in $\partial_{A_\gamma} \sigma_{\alpha\beta}^{(\bA)}$ the Drude and Berry curvature derivatives, which vanish in \T{} and \PT{} symmetric parity-mixed superconductors, respectively.
From Eq.~\eqref{anomalous_photocurrent_Green_function_method}, we see that the nonreciprocal optical responses such as the photocurrent creation ($\omega=0,~\Omega\neq 0$) and the second harmonic generation ($\omega = 2\Omega$) show a prominent divergent behavior in the low-frequency regime, which is unique to superconductors~\cite{WatanabeDaidoYanase}.
 
Recalling the minimal coupling between the electrons and electromagnetic field, the vector potential twists the phase of superfluid and plays the same role as the supercurrent. Thus, when the electromagnetic perturbation is weak, the NRSF determines the nonreciprocal component of the supercurrent induced by a given phase twist.
In the Josephson junction, the phase twist is similarly accumulated through the junction bridging the superconducting leads, and the NRSF also participates in the nonreciprocal Josephson current~\cite{baumgartner2021josephson,he2021phenomenological}.

According to the Ginzburg-Landau free energy analysis, both the NRSF and the nonreciprocal critical current are attributed to the cubic gradient component of the quadratic term as well as the linear gradient component of the quartic term~\cite{daido2021intrinsic,supple}.
Therefore, the NRSF provides a systematic understanding of various nonreciprocal responses in superconductors, including the optical response, Josephson effect, and critical current.
This is similar to the case of the conventional superfluid density, which determines the anomalous linear optical conductivity, Meissner effect, and zero-resistance phenomenon~\cite{Tinkham1959SCskin,Tinkham2004introduction}.

\textit{Nonreciprocal Meissner effect ---}
We now transform the conductivity into the susceptibility
		\begin{equation}
                J_\alpha (\omega)= K^{(1)}_{\alpha\beta} A_{\beta} (\omega) + \int \frac{d\Omega}{2\pi} K^{(2)}_{\alpha;\beta\gamma} (\omega,\Omega) A_{\beta} (\Omega)A_{\gamma} (\omega-\Omega).\label{nonlinear_Meissner_response_formula}
                \end{equation}
The anomalous nonreciprocal conductivity contributes to the response function $K^{(2)}$ as
		\begin{align}
               &2K^{(2)}_{\alpha;\beta\gamma} (\omega,\Omega)
                        = -2 \Omega \left( \omega -\Omega \right) \sigma_{\alpha;\beta\gamma}(\omega,\Omega),\\
                        &=- f_{\alpha\beta \gamma}  +\frac{i}{2}  \limvec \left( \Omega \, \partial_{A_\beta} \sigma_{\alpha\gamma}^{(\bA)} + \left( \omega-\Omega \right) \partial_{A_\gamma} \sigma_{\alpha\beta}^{(\bA)}   \right).
                \end{align}
The anomalous conductivity determines low-frequency behaviors of the nonlinear coupling between the vector potential and electric current.
This indicates that the NRSF causes the nonreciprocal property of the Meissner response, that is, nonreciprocal Meissner effect. Note that the nonreciprocal Meissner effect is unique to the parity-breaking superconductors and distinguished from the \textit{nonlinear} Meissner response~\cite{Yip1992,Xu1995Nonlinear}, which is reciprocal in terms of magnetic fields. Although the conductivity derivative may participate in the AC nonreciprocal Meissner response, we hereafter focus on the static response determined by the NRSF. Classification of the nonreciprocal Meissner kernel $K^{(2)}$ based on the \T{} and \PT{} symmetries is summarized in Table~\ref{Table_nonreciprocal_meissner_classification}.

		\begin{table}[htbp]
                \caption{Classification of the nonreciprocal Meissner kernel $K^{(2)}$ based on \T{} and \PT{} symmetries.
                The $O(\Omega^n)$ contributions allowed by the symmetry are summarized.
                }
                \label{Table_nonreciprocal_meissner_classification}
                \centering
                \begin{tabular}{ccc}\hline\hline
                        $O(\Omega^n)$ &\T{} & \PT{}\\ \hline
                        $n=0$ & N/A & NRSF\\
                        $n=1$ & Berry curvature deriv. & Drude deriv. \\ 
                        $(n\geq 2)$ & \multicolumn{2}{c}{(regularized nonlinear conductivity)} \\ 
                        \hline\hline
                      \end{tabular}
                \end{table}

We phenomenologically introduce a nonlinear correction to the London theory by
                \begin{equation}
                J_{\alpha}(\bm{r}) = - \rho_{\alpha\beta}^s A_{\beta}(\bm{r}) -f_{\alpha\beta\gamma} A_{\beta}(\bm{r})A_{\gamma}(\bm{r}),
                \end{equation}
where the first and second terms are normal and nonreciprocal supercurrents, respectively. Here, we consider a superconductor occupying the spatial region $z\leq 0$ and the NRSF vector $\bm{f} \parallel \hat{x}$ (Fig.~\ref{Fig_schematic}). When the magnetic field is applied to the $y$-direction and the nonreciprocal effect is assumed to be small, the field-dependent magnetic penetration depth is estimated as $\lambda (B) = \lambda_\text{L} \left(1 + \lambda_\text{L} B f/3 \rho^\text{s} \right)$ with the London penetration depth $\lambda_\text{L}^{-1} =\sqrt{ \mu_0\rho^\text{s} } $. Thus, it shows the unidirectional magnetic-field dependence.
Intuitively, the magnetic flux is retracted from or drawn into the superconductor when the supercurrent shielding the magnetic flux is parallel or antiparallel to the NRSF vector $\bm{f}$. Thus, a careful magnetic penetration depth measurement~\cite{Hashimoto2009penetration} can evaluate the NRSF.

The \Pa{} and \T{} symmetries have to be broken in superconductors that host the NRSF.
To our best knowledge, three setups are available; (i) systems in which \Pa{} and \T{} symmetries are broken by other spontaneous orders or by the crystal structure, (ii) superconductors under the supercurrent flow, and (iii) exotic superconductors whose order parameter spontaneously breaks the symmetries.
Case (i) is realized in various situations, such as the noncentrosymmetric superconductors under external magnetic fields~\cite{Ideue2021} and the superconductors undergoing the magnetically parity-breaking order~\cite{Sumita2017}.
Interestingly, Case (ii) was recently supported by an experiment where the superconducting NbN thin film was probed under the electric current by the second harmonic generation~\cite{Nakamura2020}.
Case (iii) is further classified into two classes. First, the multiple transitions of even-parity and odd-parity superconductivity make both \Pa{} and \T{} parities ill-defined~\cite{Wang2017}. %are opposite 
Second, these symmetries are broken by chiral superconductivity and noncentrosymmetric crystal structure~\cite{Brydon2019,Ghosh2020}.
Later we will investigate the former class in (iii) by referring to the recent proposal for a heavy fermion superconductor \utt{}~\cite{Ishizuka2021}.

We discuss the magnitude of the nonreciprocal Meissner response by the ratio
        \begin{equation}
        \eta_\text{NR} = \frac{\lambda_\text{L} B_\text{c2} f}{\rho^\text{s}}, \label{nonreciprocal_penetration_depth_ratio}
        \end{equation}
where $B_\text{c2}$ is the upper critical magnetic field. First, our analysis based on the Ginzburg-Landau theory shows $\eta_\text{NR} \propto |T-T_\text{c}|^{1/2}$~\cite{supple}, and thus, the nonreciprocal response may be negligible in the vicinity of the transition temperature $T\lesssim T_\text{c}$.
Next, we estimate the renormalization effect on the ratio $\eta_\text{NR}$. Interestingly, the correlation-induced renormalization effect denoted by $z$ positively influences the nonreciprocal Meissner response. Since $\eta_\text{NR} \propto z^{-5/2}$, the nonreciprocal Meissner effect is significantly enhanced by a strong renormalization. Thus, strongly-correlated electron systems are potential candidates offering a sizable nonreciprocal Meissner effect. While we will work on a heavy fermion system \utt{} in the following, another strongly correlated electron systems such as cuprate superconductors and twisted bilayer graphene are also of interest. The cuprates are usually centrosymmetric in the bulk, whereas the parity violation can be evoked by spontaneous order [Case (i)] and by the supercurrent injection [Case (ii)]. As for the former case, the loop-current order has been proposed for the pseudogap phase in cuprate superconductors~\cite{Varma1997,Fauque2006,Li2008-hj,Pershoguba2013-jy,Pershoguba2014-dh,Zhao2017-ka,Lim2020}. 
The NRSF may be a long-sought probe for examining such intertwining order in cuprates.

\textit{Microscopic calculations of \utt{} model---}
\utt{} is recently attracting a lot of attention as a candidate material for spin-triplet superconductivity~\cite{Ran2019,Aoki2019}. It is argued that the ferromagnetic fluctuation plays a key role in the spin-triplet superconductivity as in ferromagnetic superconductors~\cite{Ran2019,Aoki2019,Aoki_review2019,Tokunaga2019UTe2NMR,Sundar2019}, whereas the antiferromagnetic fluctuation has also been observed recently~\cite{Thomas2020,Duan2020,knafo2021lowdimensional,duan2021resonance}. 
Since the antiferromagnetic fluctuation usually stabilizes even-parity spin-singlet superconductivity, multiple magnetic fluctuations are expected to lead to multiple pairing instabilities~\cite{Ishizuka2021}. Interestingly, \utt{} shows the multiple superconducting transitions~\cite{Braithwaite2019}. Thus, the coexisting even- and odd-parity pairing state has been proposed for the low-temperature phase~\cite{Ishizuka2021}.

Based on a microscopic model, we investigate the NRSF in the putative parity-mixed phase of \utt{}. 
The model Hamiltonian for the normal state reads
        \begin{equation}
                H_{\bk}^\text{N} = \left( \varepsilon_0- \mu  \right) + V \rho_x + V' \rho_y   + \bm{g}\cdot \bm{\sigma} \rho_z,
        \end{equation}
which is described by Pauli matrices representing spin ($\sigma_\mu$) and sublattice ($\rho_\mu$) degrees of freedom. The details are given in Supplemental Materials~\cite{supple}. The model Hamiltonian reproduces the heavy band mainly consisting of U $5f$-orbitals near the Fermi level, which was obtained in the DFT+U calculations~\cite{Ishizuka2019,Shishidou2021}. We introduce a pair potential for parity-mixed superconductivity
        \begin{equation}
        \hat{\Delta}_\bk =  \left( \psi_\bk  + \bm{d}_\bk \cdot \bm{\sigma} \right)i\sigma_y \rho_0, 
        \end{equation}
where we consider intra-sublattice Cooper pairing. According to theoretical calculations implementing the Eliashberg theory and the DFT+U calculation, the even-parity pairing is characterized by the $A_g$ irreducible representation ($\psi_\bk = \Delta_\text{e} \cos{k_x}$), while the odd-parity pairing is either of the $A_u$- or $B_{3u}$-type denoted by $\bm{d}_\bk = \Delta_\text{o} \sin{k_y} \hat{y}$ or $\Delta_\text{o} \sin{k_y} \hat{z}$, respectively~\cite{Ishizuka2021}. It is energetically favorable for the relative phase between the pair potentials to be $\pm \pi/2$, when the spin-orbit coupling due to noncentrosymmetric crystal structures is absent or weak. This choice leads to the $s+ip$-wave superconductivity preserving the \PT{} symmetry~\cite{Wang2017,KanasugiYanase}.
This case contrasts with the fact that the spin-orbit coupling in the noncentrosymmetric superconductor leads to the zero phase difference indicating the \T{} symmetric state such as $s+p$-wave superconductivity~\cite{NCSC_book}.
Since \utt{} crystallizes in a centrosymmetric structure, we take the \PT{} symmetric mean field $\Delta_\text{e}^{(0)} = r \Delta_\text{e+o}^{(0)},~\Delta_\text{o}^{(0)} = i (1-r)\Delta_\text{e+o}^{(0)}$ with the parity-mixing ratio $r$. Here we denote pair potentials at zero temperature by those with the superscript `$(0)$'.

                \begin{figure}[htbp]
                \centering
                \includegraphics[width=0.98\linewidth,clip]{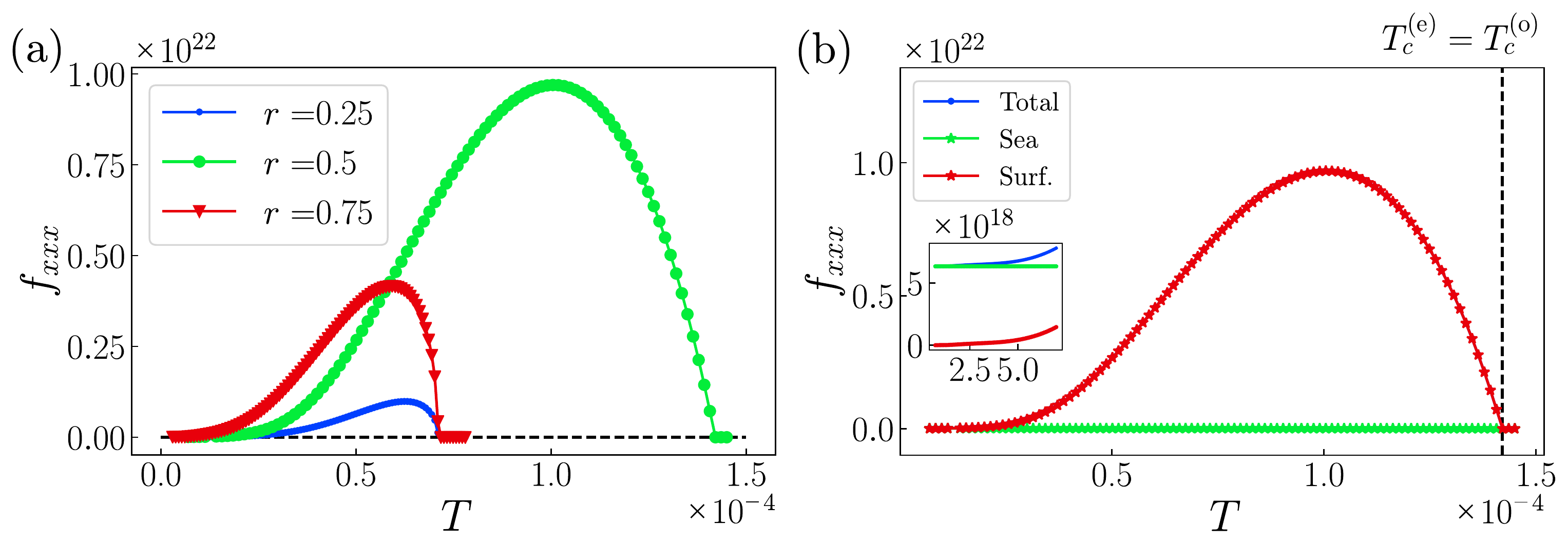}
                \caption{Temperature dependence of the NRSF $f_{xxx}$ for the $A_g + iB_{3u}$ superconducting state. (a) Plot with several ratios of even- to odd-parity pair potentials, $r=0.25$, $0.5$, and $0.75$. (b) Decomposition the total NRSF into the Fermi-sea and Fermi-surface terms in the case of $r=0.5$. The dashed line guides the transition temperature $T_\text{c}^\text{(e)} =T_\text{c}^\text{(o)} =0.5  \Delta_\text{e+o}^{(0)}/1.76 $. The inset shows the low-temperature regime with the horizontal axis $T\times 10^6$.} 
                \label{Fig_UTe3_NRSF}
                \end{figure}

Following the symmetry analysis, we obtain the NRSF $f_{xyz}$ for the $A_g + i A_u$ state and $f_{xxx},f_{xyy},f_{xzz}$ for the $A_g + i B_{3u}$ state. Here, we investigate the temperature and parity-mixing ratio dependence of the NRSF $f_{xxx}$ in the $A_g + iB_{3u}$ state in details, while we obtain a similar result for the $A_g + i A_u $ state~\cite{supple}.
The temperature dependence of the pairing potential is assumed to follow the phenomenological formula
        \begin{equation}
          \Delta_\text{e,o} (T) =  \Delta_\text{e,o}^{(0)} \tanh{\left(1.74\sqrt{\dfrac{|\Delta_\text{e,o}^{(0)}|}{1.76 T}-1}\right)}.
        \end{equation}
Figure~\ref{Fig_UTe3_NRSF} shows the NRSF calculated with several parity-mixing ratios $r=0.25$, $0.5$, $0.75$. We do not have any NRSF in the pure spin-singlet or spin-triplet state ($r=1,0$) where the \Pa{} and U(1)$\times$\Pa{} symmetry respectively forbid the NRSF. In Fig.~\ref{Fig_UTe3_NRSF}, it is clearly shown that the NRSF arises in the parity-mixed superconducting state. Each plot shows the maximum value at an intermediate temperature. This is because the NRSF is almost determined by the Fermi-surface contribution [Fig.~\ref{Fig_UTe3_NRSF}(b)] given by
        \begin{equation}
        \sum_a \left(\vj_{aa}^x  \right)^3 \frac{\partial^2 f (\varepsilon)}{\partial \varepsilon^2}_{|\varepsilon= \varepsilon_a},
        \end{equation} 
where $\bm{\vj}_{aa}$ is the paramagnetic current density of the Bogoliubov quasiparticle labeled by the quantum number $a$.
On the other hand, the Fermi-surface term gets suppressed at low temperature, and then the Fermi-sea term mainly contributes to the NRSF [inset of Fig.~\ref{Fig_UTe3_NRSF}(b)]. Decomposition of the NRSF into the Fermi-surface and Fermi-sea terms is formulated in Supplemental Materials~\cite{supple}. Since the Fermi-surface contribution is much larger than the Fermi-sea contribution, the NRSF shows a non-monotonic temperature dependence.
The sizable Fermi-surface contribution is attributed to the almost nodal superconducting gap. To support this argument, we show that the Fermi-surface term is negligible in a superconducting state with a nearly isotropic gap~\cite{supple}. Therefore, the significant NRSF and its non-monotonic temperature dependence are characteristic behaviors of nodal superconductors, and they are
useful in identifying the nodal texture in the superconducting gap.

The present study clarified that the Fermi-surface effect is much more significant than the Fermi-sea effect.
The behavior is in contrast to the normal superfluid density, which is usually determined by the Fermi-sea effect and detrimentally influenced by the Fermi-surface effect. Since the quasiparticle excitations moderately occur in the intermediate temperature regime, the NRSF $f_{\alpha\beta\gamma}$ as well as the ratio $\eta_\text{NR}$ in Eq.~\eqref{nonreciprocal_penetration_depth_ratio} are enhanced there. 

To estimate the ratio $\eta_\text{NR}$ of our model,
we first take $\rho_{xx}^\text{s} \sim 10^{19}$\,A$\cdot$V$^{-1}\cdot$m$^{-1}\cdot$s$^{-1}$~\cite{supple}, $|f_{xxx}| \sim 10^{22}$\,A$^2\cdot$V$^{-2}\cdot$s$^{-2}$, and $B_\text{c2}\sim 1 \text{\,T}$. The superfluid density leads to the penetration depth $\lambda_\text{L} \sim \mr{0.3}{\text{\textmu}m}$. Then, we obtain $\eta_\text{NR} \sim 3\times 10^{-4}$.
The ratio may increase due to the renormalization effect by electron correlations. The adopted model Hamiltonian is based on the DFT+U calculation and does not sufficiently take into account the electron correlation effect of \utt{}. We consider the renormalization factor $z=0.1$, which enhances the London penetration depth as much as the observed values $\lambda_\text{L} \sim \mr{1}{\text{\textmu}m}$~\cite{Sundar2019,Metz2019,Bae2021}. Accordingly, the ratio $\eta_\text{NR}$ increases by $\sim 300$ times larger than the above estimation. As a result, the enhanced ratio is estimated to be $\eta_\text{NR}\sim 10^{-1}$ which may be within the experimental sensitivity.

\textit{Discussion and Summary---}
This Letter reveals that the NRSF plays an essential role in various nonreciprocal responses of superconductors such as optical responses, supercurrent flow, and Meissner effect. Thus, the NRSF is a potential indicator of nonreciprocal responses and is helpful to probe the parity mixing in a superconducting state, which causes parity breaking required for the nonreciprocal superconducting phenomena. It is noteworthy that the NRSF can be estimated through various experimental techniques implemented in nonlinear optics and magnetic penetration depth measurements. We also note that the NRSF is well-defined in the whole temperature region and complementary to the fluctuation-assisted normal nonlinear conductivity, which captures the nonreciprocal response in the vicinity of the transition temperature~\cite{Wakatsuki2018,Wakatsuki2017,Hoshino2018}.

We have mainly discussed the NRSF and the resulting nonreciprocal Meissner response in exotic superconductors, where the competing pairing interaction in the spin-singlet and spin-triplet channels leads to multiple superconducting transitions. However, our formulation applies to the NRSF in a broad class of superconductors with broken \Pa{} and \T{} symmetries. For instance, the noncentrosymmetric superconductors under the external magnetic field and some classes of magnetic superconductors satisfy the symmetry condition. As artificially-engineered noncentrosymmetric superconductors have realized sizable nonreciprocal transport phenomena in recent experimental works~\cite{Ando2020,baumgartner2021josephson}, it is also expected that nonreciprocal properties of optical phenomena and Meissner response due to the NRSF are observed.

The dynamical nonreciprocal response is also of interest, while we mainly focused on the static nonreciprocal Meissner response in this study. The low-frequency behavior of the nonreciprocal Meissner Kernel $K^{(2)}$ is closely related to the space-time symmetry of the superconductivity as shown in Table~\ref{Table_nonreciprocal_meissner_classification}; $K^{(2)}$ has a static component in the \PT{} symmetric parity-mixed superconductor, whereas it contains a $\Omega$ linear term as a leading order term in the \T{} symmetric superconductor. Thus, a careful AC experiment on the magnetic penetration depth may distinguish the symmetry of parity-mixed superconductors.

To summarize, we proposed that the NRSF provides a systematic understanding of nonreciprocal responses in superconductors. Accordingly, we clarified the nonreciprocal magnetic penetration phenomenon, which we call the nonreciprocal Meissner effect. According to the study of \utt{}, the nonreciprocal Meissner effect mainly arises from the Fermi-surface effect and is sensitive to the superconducting gap structure and electron correlation effect. Therefore, the phenomenon is expected to be a key to identify the symmetry of exotic superconductors such as \utt{}.

Supplemental Materials include Refs.~\cite{Ikeda2006UTe2crystalstructure,hutanu2019crystal,Fischer2011,Maruyama2012-pu,Basov2005-pg,Ahn2021SC,Watanabe2020,Zelezny2014,ishihara2021chiral}

\section*{Acknowledgement}
The authors thank Jun Ishizuka, Shota Kanasugi, and Yoshihiro Michishita for fruitful discussions.
This work was supported by JSPS KAKENHI (Grants No. JP18H05227, No. JP18H01178, and No. 20H05159) and SPIRITS 2020 of Kyoto University.
H.W. is a JSPS research fellow and supported by JSPS KAKENHI (Grant No.~18J23115 and No.~21J00453).
A.D. is supported by JSPS KAKENHI (Grant No.~21K13880).

\clearpage
\onecolumngrid 

\renewcommand{\thesection}{S\arabic{section}}
\renewcommand{\theequation}{S\arabic{equation}}
\setcounter{equation}{0}
\renewcommand{\thefigure}{S\arabic{figure}}
\setcounter{figure}{0}
\renewcommand{\thetable}{S\arabic{table}}
\setcounter{table}{0}
\makeatletter
\c@secnumdepth = 2
\makeatother

\begin{center}
\vspace{1cm}
\textbf{\large Supplemental Materials for\\ ``Nonreciprocal Meissner Response in Parity-Mixed Superconductors''}\\
\vspace{0.5cm}
Hikaru Watanabe, Akito Daido, and Youichi Yanase
\vspace{0.5cm}
\end{center}

\section{Ginzburg-Landau analysis of nonreciprocal superfluid density}
\label{App_Sec_GL}

In this section, we estimate the NRSF based on the Ginzburg-Landau (GL) analysis. As mentioned in the main text, the NRSF appears in three-fold cases; (I) superconductors where \Pa{} and \T{} symmetries are broken in the normal state, (II) superconductors under the supercurrent, and (III) the cases of superconductivity which itself breaks the symmetry. Most cases are captured by the single component GL free energy, while the multicomponent GL theory should be analyzed when the multiple superconducting transitions break the symmetry as in Case (III). We, therefore, present two GL analyses dealing with a single component or multicomponent order parameter.

\subsection{Single component GL theory}

We consider the GL free energy parametrized by the single component order parameter $\Delta$. The GL free energy is given by~\cite{daido2021intrinsic}
    \begin{equation}
    F[q,\Delta] = \alpha (q) \Delta^2 + \frac{1}{2} \beta (q) \Delta^4,    \end{equation}
where the quadratic and quartic terms read
	\begin{align}
    &\alpha (q) = -a_0 + \frac{1}{2}a_2 (q-q_0)^2+ \frac{1}{6}a_3 (q-q_0)^3,\\
    &\beta (q) = b_0 + b_1 (q-q_0).
    \end{align}
The gradient terms are obtained by the expansion of the GL free energy around $q_0$.
For instance, a finite $q_0$ arises from the cooperation of an antisymmetric spin-orbit coupling and an external magnetic field [Case (I)] or from an injected supercurrent [Case (II)].
We also impose the conditions, $\alpha (q)<0$ and $\beta (q)>0$, to obtain a non-zero proper solution for the order parameter.

With the saddle-point approximation for the order parameter, $\partial_{\Delta} F[q,\Delta] = 0$, the order parameter is given by
	\begin{equation}
    \overline{\Delta}^2 = - \frac{\alpha (q)}{\beta (q)}.\label{App_saddle_point_approximation}
    \end{equation}
Plugging the obtained order parameter into the GL free energy, we obtain  
		\begin{equation}
    \mathcal{F} (q) = F[q,\overline{\Delta}] = -\frac{1}{2} \frac{\alpha^2 (q)}{\beta (q)}.
    \end{equation}
The NRSF $f_\text{NRSF}$ is obtained from the third derivative with respect to $q$ as
	\begin{equation}
    f_\text{NRSF} = \lim_{q\rightarrow 0}  \partial_q^3 \mathcal{F} (q) =  a_0 \left( \frac{1}{b_0}a_3  -\frac{3a_2}{b_0^2}b_1 \right) + O(b_1^3).
    \end{equation}
As in the cases of the nonreciprocal critical current and nonreciprocal Josephson effect~\cite{yuan2021supercurrent,daido2021intrinsic,he2021phenomenological}, higher-order gradient terms denoted by $a_3, b_1$ play essential roles in the nonreciprocal property of the superfluid density. When we consider the temperature dependence of the mass term $a_0 \propto |T_\text{c}-T|$, it follows that $f_\text{NRSF} \propto |T_\text{c}-T|$.
Taking into account the temperature dependence of the upper critical field and magnetic penetration depth, we obtain the critical behavior, $\eta_\text{NR} \propto |T-T_{\rm c}|^{1/2}$.

\subsection{Multicomponent GL theory}
Here we consider 
the GL free energy consisting of two component order parameters $\Delta_1,\,\Delta_2$ whose \Pa{} parities are opposite. We assume the relative phase between the two order parameters to be $\pm \pi/2$ and redefine the order parameters by $(\Delta_1,\Delta_2)\rightarrow (\Delta_1, i\Delta_2)$ to suppress the imaginary unit. The GL free energy is given by
	\begin{equation}
    F[q_1,q_2,\Delta_1,\Delta_2] = \alpha_1 (q_1) \Delta_1^2 + \alpha_2 (q_2) \Delta_2^2 + \frac{1}{2} \beta_1 (q_1) \Delta_1^4 + \frac{1}{2} \beta_2 (q_2) \Delta_2^4 + F_{12}.
    \end{equation}
The coupling coefficients of the quadratic and quartic terms are defined as
	\begin{align}
    &\alpha_i (q_i ) = -a_0^{(i)} + \frac{1}{2}a_2^{(i)} q_i^2, \label{app_quadratic_coefficient_double_GL} \\
    &\beta_i (q_i ) = b_0^{(i)}.\label{app_quaratic_coefficient_double_GL}
    \end{align}
Since the normal state is assumed to be centrosymmetric, we have no components arising from the parity violation in Eqs.~\eqref{app_quadratic_coefficient_double_GL} and~\eqref{app_quaratic_coefficient_double_GL}. The cross-coupling term between $\Delta_1$ and $\Delta_2$ reads
	\begin{equation}
    F_{12} =  \frac{c_3}{12} (q_1^3 + q_2^3) \Delta_1  \Delta_2  + \frac{1}{2} d_0  \Delta_1^2 \Delta_2^2.\label{eq:F12_def}
    \end{equation}
The coefficient $c_3$ is allowed in the \Pa{}-broken system and thus characteristic to multi-component  superconducting order parameters which cooperatively break the \Pa{} symmetry. Note that we assume the parity-mixed superconducting state which is nonpolar but noncentrosymmetric for simplicity. 
Here, we consider two cases; (1) $\Delta_1$ and $\Delta_2$ have the same transition temperature and (2) the transition temperature for $\Delta_1$ is much lower than that for $\Delta_2$. In the latter case, we can neglect the temperature dependence of $\Delta_2$.
In general, the transition temperatures for $\Delta_1$ and $\Delta_2$ are not equivalent without fine tuning.

First, we consider the case (1) where the mass terms $a_0^{(i)}$ vanish at the same temperature. Regarding $c_3$ as a small parameter, we apply the saddle-point approximation to the order parameters. The order parameters obtained with neglecting $c_3$ are given by
        \begin{equation}
            \begin{pmatrix}
            {\overline{\Delta}_1}^2\\
            {\overline{\Delta}_2}^2
            \end{pmatrix}
            = \frac{-2}{4b_0^{(1)} b_0^{(2)} -d_0^2}
            \begin{pmatrix}
            2b_0^{(2)}\alpha_1 (q_1)-d_0 \alpha_2 (q_2)\\
            -d_0 \alpha_1 (q_1) +2 b_0^{(1)} \alpha_2 (q_2)
            \end{pmatrix}.
        \end{equation}
Plugging $\overline{\Delta}_1$ and $\overline{\Delta}_2$ into the first term of Eq.~\eqref{eq:F12_def}, we obtain $f_{\text{NRSF}}$ up to first order in $c_3$,
	\begin{align}
    f_\text{NRSF} 
        &= \lim_{q_1,q_2 \rightarrow 0} \sum_{0\leq n,m }^{n+m=3} \frac{(n+m)!}{n!~ m!} ~ \partial_{q_1}^n\partial_{q_2}^m F [q_1,q_2,\overline{\Delta_1},\overline{\Delta_2}],\\
        &= \frac{2c_3}{ 4b_0^{(1)} b_0^{(2)}-d_{0}^2}  \sqrt{\left( a_0^{(2)} d_{0}-2 a_0^{(1)} b_0^{(2)}\right)  \left( a_0^{(1)} d_{0}-2 a_0^{(2)} b_0^{(1)} \right)}.
    \end{align}
Taking into account the temperature dependence of the mass terms $a_0^{(i)}\propto |T_\text{c}-T|$, we obtain $f_\text{NRSF} \propto |T_\text{c}-T|$.
As a result, the critical behavior of the NRSF is the same as that in the single component GL theory.  
 
Next, we consider the case (2). Neglecting the temperature dependence of an order parameter $\Delta_2$, we discuss the second superconducting transition due to the appearance of $\Delta_1$. 
Thus, the following calculation is based on the GL free energy for $\Delta_1$
	\begin{equation}
    F' [q,\Delta_1] = \alpha' (q) \Delta_1^2  + \frac{1}{2} \beta' (q) \Delta_1^4 + \frac{1}{6} \kappa_3 q^3 \Delta_1,
    \end{equation}
where we give the coefficients by
	\begin{align}
    &\alpha' (q) = -A_0 + \frac{1}{2}A_2 q^2,\\
    &\beta' (q) = B_0.
    \end{align}
The $O(\Delta_1)$ term denoted by $\kappa_3$ originates from the coupling with the order parameter $\Delta_2$, although the odd-order terms $\Delta^{2n+1}$ of the GL free energy are usually forbidden due to the global gauge symmetry. If the coupling constant $\kappa_3$ is sufficiently small, the saddle-point approximation can be performed as in Eq.~\eqref{App_saddle_point_approximation}. Accordingly, the GL free energy is obtained as
        \begin{equation}
            \mathcal{F}' (q) = F' [q,\overline{\Delta}_1],
        \end{equation}
and the NRSF is evaluated up to $O(\kappa_3)$ by
	\begin{equation}
    f_\text{NRSF} = \lim_{q\rightarrow 0}  \partial_q^3 \mathcal{F}' (q)= \kappa_3 \sqrt{\frac{A_0}{B_0}}.
    \end{equation}
The temperature dependence of the NRSF is $|T_\text{c}-T|^{1/2}$ due to $A_0 \propto |T_\text{c}-T|$. Note that the normal superfluid density is less dependent on $T_\text{c}-T$ since the order parameter $\Delta_2$ makes a dominant contribution to the superfluidity in the critical region of the second superconducting transition.

To summarize, the critical exponents of the NRSF in multi-component superconductors are different between the equally pairing case ($r=0.5$) and the unequally pairing case ($r\neq 0.5$). The critical behaviors of the NRSF ($f_\text{NRSF} \propto |T_\text{c} -T|^\gamma$, $\gamma=1, 0.5$) are confirmed by the numerical analysis of the model Hamiltonian adopted in the main text.
In both cases, we obtain the critical exponent for $\eta_\text{NR} \propto |T-T_\text{c}|^{1/2}$.

\section{Model study for Uranium Ditelluride}
\label{App_Sec_UTe2_model}

\begin{figure}[htbp]
                        \centering
                        \includegraphics[width=0.75\linewidth,clip]{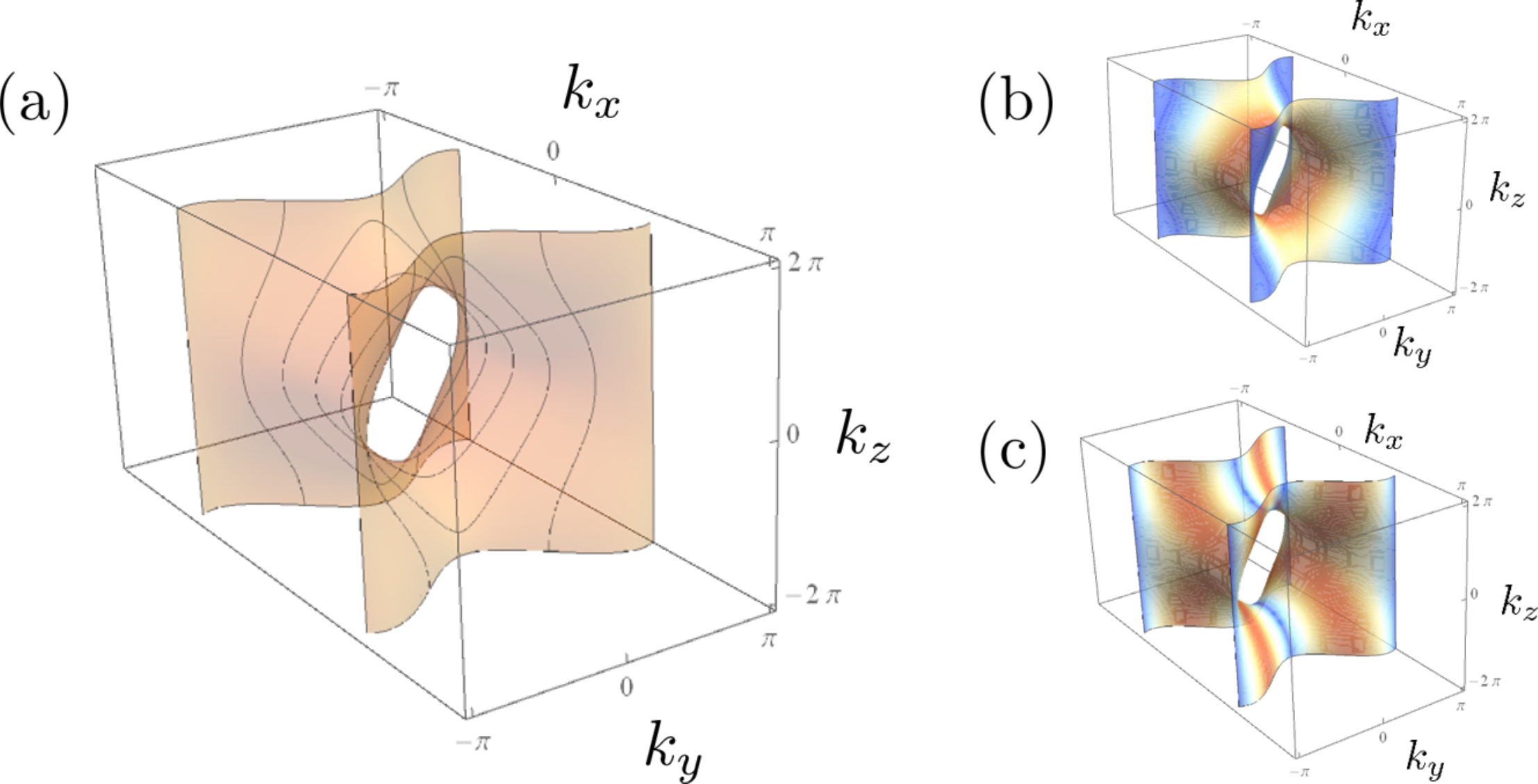}
                        \caption{(a) Fermi surface of the effective single orbital model for \utt{}. (b) and (c) Momentum dependence of the gap amplitude for (b) the even-parity pair potential and (c) the odd-parity pair potential.}
                    \label{App_Fig_UTT_FS_gap}
                    \end{figure}

We introduce the model Hamiltonian proposed in Ref.~\cite{Shishidou2021} and investigate the superconductivity of \utt{}. \utt{} has the orthorhombic crystal structure (space group $Immm$, No.~71) and U sites show the locally-noncentrosymmetric property. The model Hamiltonian for the normal state is given by  
        \begin{equation}
        H_{\bk}^\text{N} = \left( \varepsilon_0- \mu  \right) + V \rho_x+ V' \rho_y + \bm{g}\cdot \bm{\sigma} \rho_z,\label{App_UTe2_normal_hamiltonian}
        \end{equation}
where the spin and sublattice degrees of freedom are respectively represented by the Pauli matrices $\bm{\sigma}$ and $\bm{\rho}$. Each component is given by
        \begin{align}
        \varepsilon_0 &=t_{1} \cos{k_{x}}+t_{2} \cos{k_{y}}, \\
        V &=\left( v_1 + v_2 \cos{\frac{k_x}{2}}\cos{\frac{k_y}{2}} \right) \cos{\left( 2k_z \delta - \frac{k_z}{2} \right) }, \\
        V' &=\left( v_1 + v_2 \cos{\frac{k_x}{2}}\cos{\frac{k_y}{2}}  \right) \sin{\left( 2k_z \delta - \frac{k_z}{2} \right) }, \\
        \bm{g} &= \left(  \alpha_1 \sin{k_y},\alpha_2 \sin{k_x},\alpha_3 \sin{\frac{k_{x}}{2}} \sin{\frac{k_{y}}{2}} \sin{\frac{k_{z}}{2}} \right),
        \end{align}
with $\delta = 0.135$~\cite{Ikeda2006UTe2crystalstructure,hutanu2019crystal}. The locally noncentrosymmetric property is built into the sublattice-dependent antisymmetric spin-orbit coupling $\bm{g}\cdot \bm{\sigma} \rho_z$~\cite{Fischer2011,Maruyama2012-pu}. The model parameters are determined to reproduce the Fermi-surface consisting of U $5f$ orbitals~\cite{Shishidou2021,Ishizuka2019}. The parameters in the unit of electron volt (eV) are given by 
        \begin{equation}
        (\mu,t_1,t_2,v_1,v_2,\alpha_1,\alpha_2,\alpha_3)=(-0.129,-0.0892,0.0678,-0.062,0.0742,0.006,0.008,0.01).
        \end{equation}
Figure~\ref{App_Fig_UTT_FS_gap}(a) shows the Fermi surface. The obtained hyperbolic Fermi surface has been observed in DFT+U calculations~\cite{Ishizuka2019,Shishidou2021}.

With the normal Hamiltonian, we construct the Bogoliubov-de Gennes (BdG) Hamiltonian
    \begin{equation}
    \mathcal{H}_\text{BdG} = \frac{1}{2}\sum_\bk \bm{\Psi}_\bk^\dagger H_\text{BdG}(\bk)\bm{\Psi}_\bk,
    \end{equation}
where the $\bk$-resolved Nambu spinor is $\bm{\Psi}_\bk = \left(  \bm{c}_\bk^\dagger , \bm{c}_{-\bk}^T  \right)$. The electron creation operator $ \bm{c}_\bk^\dagger$ is labeled by the spin and sublattice degrees of freedom.
For each $\bk$, the BdG Hamiltonian is given by
    \begin{equation}
        H_\text{BdG} (\bk)
        = \begin{pmatrix}
            H_{\bk}^\text{N} & \hat{\Delta}_{\bk}\\
            \hat{\Delta}_{\bk}^\dagger& -\left( H_{-\bk}^\text{N} \right)^T
          \end{pmatrix}.\label{BdG_hamiltonian_momentum_representation}
        \end{equation}
The superconducting pair potential is given by the mixture of the odd and even-parity pair potentials
        \begin{equation}
        \hat{\Delta}_\bk =  \left( \psi_\bk  + \bm{d}_\bk \cdot \bm{\sigma} \right)i\sigma_y \rho_0, 
        \end{equation}
with $\psi_\bk =\Delta_\text{e} \cos{k_x}$ and $\bm{d}_\bk = \Delta_\text{o} \sin{k_y} \hat{y}$ ($A_u$ state) or $\bm{d}_\bk = \Delta_\text{o} \sin{k_y} \hat{z}$ ($B_{3u}$ state). Here we do not take into account the inter-sublattice pairing for simplicity.

We adopt the total magnitude of the pairing potential $\Delta_\text{e+o}^{(0)} = | \Delta_\text{e}^{(0)} | + |\Delta_\text{o}^{(0)}| = 0.0005$ at the zero temperature, where the pair potential amplitudes are parametrized by the ratio $r$ as $| \Delta_\text{e}^{(0)} |/| \Delta_\text{o}^{(0)}| = r/(1-r)$. 
The temperature dependence is assumed to be phenomenological form 
        \begin{equation}
        \Delta_\text{e,o} (T) =  \Delta_\text{e,o}^{(0)} \tanh{\left(1.74\sqrt{\dfrac{|\Delta_\text{e,o}^{(0)}|}{1.76 T}-1}\right)}.\label{App_phenomenological_SC_temperature_dependence}
        \end{equation}
The transition temperatures $T_\text{c}^\text{(e,o)} = |\Delta_\text{e,o}^{(0)}| /1.76$ roughly correspond to the observed superconducting transition temperature~\cite{Ran2019,Braithwaite2019}. We show the momentum dependence of the even and odd-parity superconducting gap amplitudes in Figs.~\ref{App_Fig_UTT_FS_gap}(b) and (c), which are parametrized by $|\cos{k_x}|$ and $|\sin{k_y}|$, respectively.

In the following subsections, we show the numerical results of all the components of normal and nonreciprocal superfluid densities, although we show a component of the NRSF $f_{xxx}$ for the $A_g + i B_{3u}$ state in the main text. 

\subsection{Normal superfluid density}
\label{App_Sec_normal_SF_ute2}

The linear optical conductivity has been calculated in the framework of the BdG formulation~\cite{Tinkham2004introduction}. The total optical conductivity is given by~\cite{WatanabeDaidoYanase} 
        \begin{equation}
        \sigma_{\alpha\beta} (\omega) = \sigma_{\alpha\beta}^\text{reg} (\omega) + \sigma_{\alpha\beta}^\text{M} (\omega). \label{total_linear_optical_conductivity}
        \end{equation}
The first component is the normal optical conductivity
	\begin{equation}
        2\sigma^\text{reg}_{\alpha\beta} (\omega) = \sum_{a} \frac{1}{i \omega-\eta} \vj_{aa}^{\alpha} \vj_{aa}^{\beta} \partial_\varepsilon f_a  + i  \sum_{a\neq b} \frac{\vj^\alpha_{ab} \vj^\beta_{ba}f_{ab}}{( \omega + i\eta -\varepsilon_{ba}) \varepsilon_{ab}},\label{regular_linear_conductivity}
        \end{equation}
which consists of the intraband and interband terms.
The energy difference $\varepsilon_{ab} = \varepsilon_{a}-\varepsilon_{b}$ and the Fermi-Dirac distribution function $f_a$ are defined with the energy eigenvalue $\varepsilon_{a}$ of the BdG Hamiltonian. 
We introduced the adiabaticity parameter $\eta$ and the current operators $\vj^{\alpha}$ and $\vj^{\alpha\beta}$ which are called paramagnetic and diamagnetic current operators, respectively.
The basis for the single particle energy eigenstates $\{\ket{a} \}$ is spanned by the crystal momentum $\bk$ and the band indices. In the Nambu representation, the paramagnetic and diamagnetic currents are given by
		\begin{equation}
                \mathcal{J}^{\alpha} = \frac{1}{2}\sum_{\bk,p,q} \Psi_{\bk p}^\dagger \vj_{pq}^{\alpha} \Psi_{\bk q},~\mathcal{J}^{\alpha\beta} = \frac{1}{2}\sum_{\bk,p,q} \Psi_{\bk p}^\dagger \vj_{pq}^{\alpha\beta} \Psi_{\bk q},  \label{App_fermi_surface_term_with_second_deriv}
                \end{equation}
where the $\bk$-resolved matrices are $\vj^\alpha =  \partial_{\alpha} H^\text{N}_\text{BdG} (\bk) \tau_z$ and $\vj^{\alpha\beta} =  \partial_{\alpha}\partial_{\beta} H^\text{N}_\text{BdG} (\bk) $ spanned by the band indices $(p,q)$. 
The Pauli matrices $\bm{\tau}$ represent the Nambu degree of freedom and $H^\text{N}_\text{BdG}$ denotes the BdG Hamiltonian without the pair potential. Although Eq.~\eqref{regular_linear_conductivity} is diverging in the low-frequency limit when the superconducting gap has a nodal structure, it is regularized by a proper scattering effect such as the electron correlation and impurity scattering. Thus, the contribution is denoted as the `regular' term~\cite{Tinkham1959SCskin,Basov2005-pg}. On the other hand, the second term in Eq.~\eqref{total_linear_optical_conductivity} represents a divergent contribution written by
        \begin{equation}
        2\sigma_{\alpha\beta}^\text{M} (\omega) = -\frac{2\rho^\text{s}_{\alpha\beta}}{i\omega - \eta}.
        \label{eq:sigmaM}
        \end{equation}
We introduced the (normal) superfluid density $\rho^\text{s}_{\alpha\beta}$
		\begin{equation}
                \rho^\text{s}_{\alpha\beta} = \limvec \partial_{A_\alpha}\partial_{A_\beta}F_A = \frac{1}{2} \limvec \sum_a \partial_{A_\beta} \left(  \partial_{A_\alpha} \varepsilon _a  f_{a}  \right),
                \end{equation}
where $F_A$ is the free energy calculated with the BdG Hamiltonian including the vector potential $\bA$. The contribution \eqref{eq:sigmaM} gives an infinite static conductivity since $\left(  w + i\eta \right)^{-1} = \text{P}\, \omega^{-1} - i\pi \delta (\omega)$ where $\text{P}$ represents the principal integral for $\omega$. This reveals that the superfluid density ensures the zero-resistivity phenomenon in superconductors. Thus, the $\omega^{-1}$ divergence in $\Im{[\,\sigma_{\alpha\beta}^\text{M} (\omega)]}$ is robust to a scattering effect unless the superfluid density vanishes.

On the basis of the London gauge ($\nabla \cdot \bA = 0$), the superfluid density gives rise to the static coupling between the electric current and vector potential, that is, Meissner response. The response formula is given by
		\begin{equation}
                J_\alpha (\bm{r}) = - \rho^\text{s} A_\alpha (\bm{r}),
                \end{equation}
for an isotropic superconductor $\rho^\text{s}_{\alpha\beta} = \rho^\text{s} \delta_{\alpha\beta}$. Accordingly, the London penetration depth is obtained as
		\begin{equation}
                \lambda_\text{L} = \frac{1}{\sqrt{\mu_0 \rho^\text{s}}},
                \end{equation}
with the vacuum permeability $\mu_0$.

Here we calculate the superfluid density in the \utt{} model. The numerical calculation is performed with the $N^3$-discretized Brillouin zone. It is convenient for numerical calculations to rewrite the superfluid density by the paramagnetic and diamagnetic current operators. The expression is given by
        \begin{align}
        \sum_a \partial_{A_\alpha } \left( \partial_{A_\beta} \varepsilon_a f_a \right) 
          &=\sum_a \partial_{A_\alpha } \left( -\vj^\beta_{aa} f_a \right),\\
          &=\sum_a \left[  \vj^{\alpha\beta}_{aa} +\sum_{b\neq a}\frac{1}{\varepsilon_{ab}} \left( \vj^\alpha_{ab}\vj^\beta_{ba} + {\rm c.c.} \right) \right] f_a +\sum_a  \vj^\alpha_{aa}\vj^\beta_{aa} \partial_\varepsilon f_a ,\label{superfluid_density_numerical}
        \end{align}
where we used the Hermann-Feynman relation between the paramagnetic and diamagnetic current operators. In the final line, the first two components are Fermi-sea contributions, whereas the last term is the Fermi-surface contribution.
When we adopt a simple model Hamiltonian such as that preserving the \Pa{} symmetry, the interband matrix element of the paramagnetic current operator is forbidden by the emergent particle-hole symmetry~\cite{Ahn2021SC}. Thus, in superconductors with gapped Bogoliubov quasiparticle spectrum and such an emergent symmetry, the superfluid density at low temperatures is determined by the diamagnetic contribution, which is closely related to the total electron density~\cite{Tinkham1959SCskin,Watanabe2020}.

\begin{figure}[htbp]
                \centering
                \includegraphics[width=0.90\linewidth,clip]{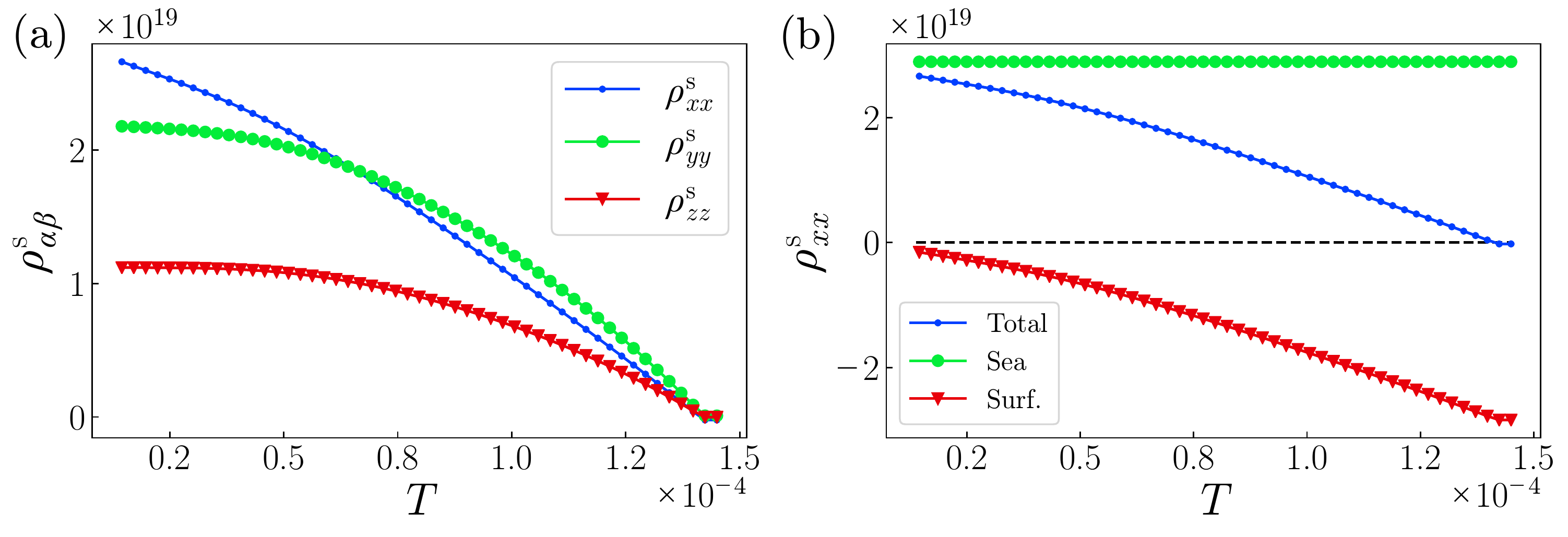}
                \caption{Superfluid density (A$\cdot$V$^{-1} \cdot$m$^{-1} \cdot$s$^{-1}$) in the $A_g + iB_{3u}$ pairing state. (a) Temperature dependence of $\rho^\text{s}_{xx}$ (blue), $\rho^\text{s}_{yy}$ (green), and $\rho^\text{s}_{zz}$ (red). (b) Decomposition of $\rho^\text{s}_{xx}$ into the Fermi-sea (green) and Fermi-surface (red) contributions. The blue-colored plot is the total superfluid density. We adopt $r=0.5$ and the lattice constants reported in Ref.~\cite{Ikeda2006UTe2crystalstructure}. We choose $N=300$ for numerical calculations.}
                \label{App_Fig_UTe2_superfluid}
                \end{figure}

Figure~\ref{App_Fig_UTe2_superfluid} shows the numerical results of the $A_g + iB_{3u}$ state with the parameter $r=0.5$.
We obtained qualitatively the same results for the $A_g + iA_{u}$ state, and another ratio for the pairing potentials $r$ gives similar results.
Owing to the orthorhombic symmetry of \utt{}, the superfluid density is anisotropic, and thus, we show $\rho_{xx}^{s}$, $\rho_{yy}^{s}$, and $\rho_{zz}^{s}$.
We also show the Fermi-sea and Fermi-surface contributions in Fig.~\ref{App_Fig_UTe2_superfluid}(b).
Although the two contributions are completely canceled in the normal state ($T\ge T_\text{c}$), a finite superfluid density appears as the Fermi-surface term is suppressed in the superconducting state ($T < T_\text{c}$). The superfluid density usually takes the maximum value determined at zero temperature, and it is determined 
by the Fermi-sea term. Thus, the Fermi-surface contribution is detrimental to the Meissner response.
The power-law temperature dependence of the superfluid density indicates anisotropic superconducting gap structure, as seen in an experiment for UTe$_2$~\cite{ishihara2021chiral}.
From Fig.~\ref{App_Fig_UTe2_superfluid} the London penetration depth is evaluated as $\lambda_\text{L} \sim \mr{0.3}{\text{\textmu}m}$ by adopting the superfluid density $\sim 10^{19}$\,A$\cdot$V$^{-1} \cdot$m$^{-1} \cdot$s$^{-1}$.

\subsection{Nonreciprocal superfluid density}
\label{App_Sec_NRSF_UTe2}

The numerical calculation of the NRSF, $f_{\alpha\beta\gamma} = \limvec \partial_{A_\alpha}  \partial_{A_\beta} \partial_{A_\gamma} F_{A}$, can be performed as in Sec.~\ref{App_Sec_normal_SF_ute2}. When we rewrite the NRSF by the velocity operators, the expression is 
		\begin{equation}
                f_{\alpha\beta\gamma} = f_ {\alpha\beta\gamma}^\text{(surf)} + f_{\alpha\beta\gamma}^\text{(sea)},
                \end{equation}
which consists of the Fermi-surface contribution
                \begin{align}
               -2 f_ {\alpha\beta\gamma}^\text{(surf)}
                &= \sum_a \left(  \vj^{\mu\nu}_{aa} \vj^{\lambda}_{aa} + \vj^{\mu\lambda}_{aa} \vj^{\nu}_{aa}+\vj^{\nu\lambda}_{aa} \vj^{\mu}_{aa} \right) \partial_\varepsilon f_a + \sum_a  \vj^\mu_{aa}\vj^\nu_{aa}\vj^\lambda_{aa} \partial^2_\varepsilon f_a \notag \\
                &+\sum_{a\neq b}  \frac{1}{\varepsilon_{ab}}\left[  \vj^\mu_{aa} \left(  \vj^{\nu}_{ab} \vj^{\lambda}_{ba} +{\rm c.c.}  \right) + \vj^\nu_{aa} \left(  \vj^{\lambda}_{ab} \vj^{\mu}_{ba} +{\rm c.c.}  \right)+\vj^\lambda_{aa} \left(  \vj^{\mu}_{ab} \vj^{\nu}_{ba} +{\rm c.c.}  \right)  \right] \partial_\varepsilon f_a,
                \label{NSF_numerical_formula_surface}
                \end{align}
and the Fermi-sea contribution
        \begin{align}
                -2f_ {\alpha\beta\gamma}^\text{(sea)}
                &=\sum_a \vj^{\alpha\beta\gamma}_{aa}f_a + \sum_{a\neq b} \frac{1}{\varepsilon_{ab}} \left( \vj^{\alpha}_{ab}\vj^{\beta\gamma}_{ba}  + \vj^{\beta}_{ab}\vj^{\gamma\alpha}_{ba} + \vj^{\gamma}_{ab}\vj^{\alpha\beta}_{ba} + {\rm c.c.}. \right)  f_a - \sum_{a\neq b}  \frac{\Delta^{\gamma}_{ab}}{\varepsilon_{ab}^2} \left( \vj^{\beta}_{ab}\vj^{\alpha}_{ba} + \vj^{\beta}_{ba}\vj^{\alpha}_{ab} \right) f_a \notag \\ 
                & + \sum_{a,b,c}^{a\neq b,c}  \frac{1}{\varepsilon_{ab}\varepsilon_{ac}} f_a \left[  \vj^\alpha_{ab}\vj^\beta_{bc}\vj^\gamma_{ca} +\vj^\beta_{ab}\vj^\alpha_{bc}\vj^\gamma_{ca} +  {\rm c.c.} \right] + \sum_{a,b,c}^{b\neq a,c} \frac{1}{\varepsilon_{ab}\varepsilon_{bc}} f_a \left[  \vj^\beta_{ab}\vj^\gamma_{bc}\vj^\alpha_{ca} + \vj^\alpha_{ab}\vj^\gamma_{bc}\vj^\beta_{ca} +  {\rm c.c.} \right].
                \label{NSF_numerical_formula_sea}
        \end{align} 
We defined the third-order generalized velocity operator, whose momentum representation is $\vj^{\alpha\beta\gamma} =  \partial_{\alpha} \partial_{\beta} \partial_{\gamma}  H^\text{N}_\text{BdG} (\bk) \tau_z$.
 The velocity difference matrix is defined by $\Delta^\alpha_{ab} = \vj^\alpha_{aa}-\vj^\alpha_{bb} = -\partial_{\alpha} \varepsilon_{ab}$.

\begin{figure}[htbp]
                \centering
                \includegraphics[width=0.50\linewidth,clip]{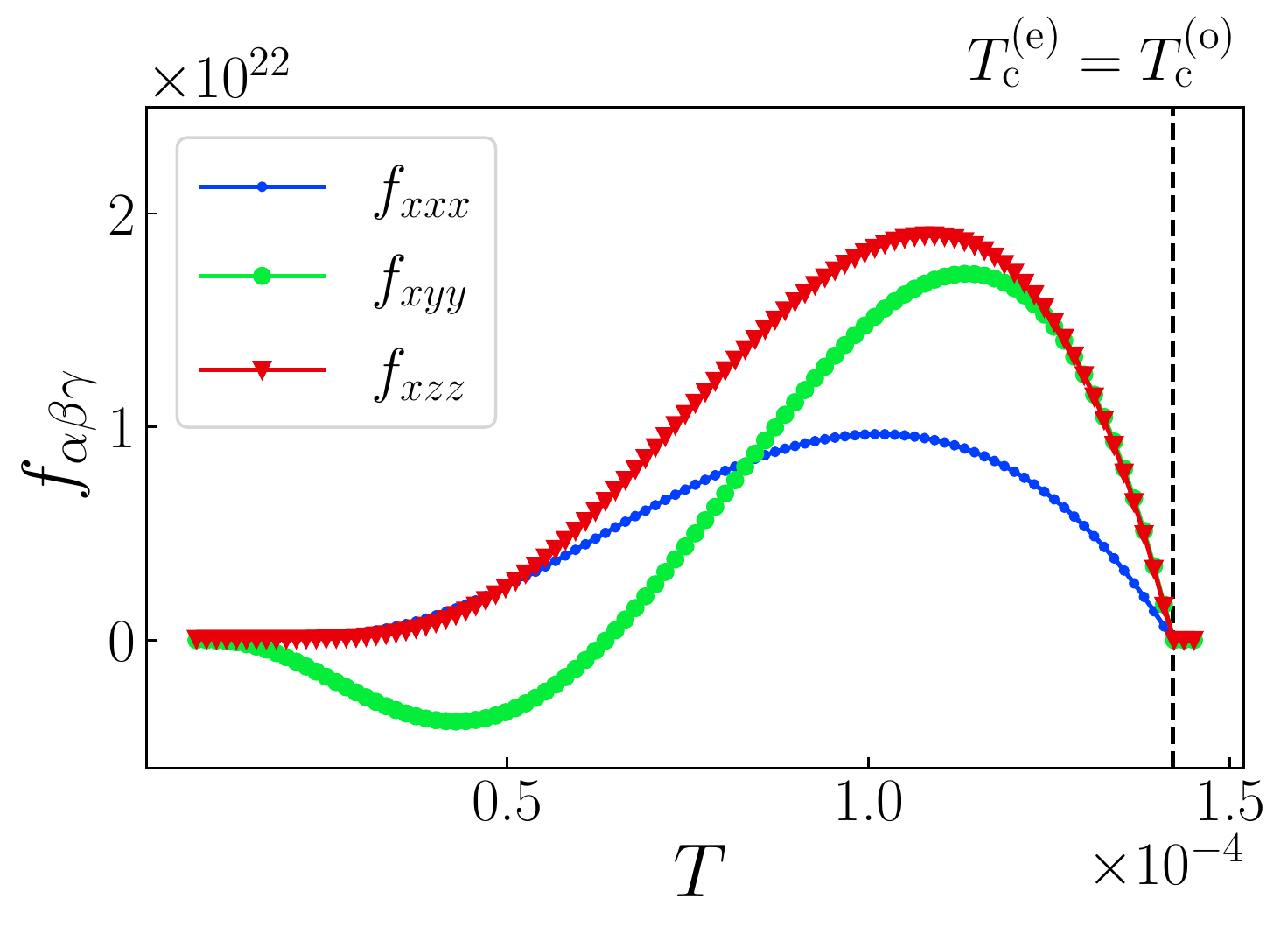}
                \caption{Temperature dependence of the NRSF tensors $f_{\alpha\beta\gamma}$ in the $A_g + iB_{3u}$ equally-pairing ($r=0.5$) state. The unit of the NRSF is A$\cdot$V$^{-2}\cdot$s$^{-2}$, and we implement the $N^3=200^3$ discretized momentum summation. The vertical dashed line guides the transition temperature.}
                \label{App_Fig_UTe2_NRSF_50_all}
                \end{figure}

First, we investigate the NRSF for the $A_g + iB_{3u}$ pairing state. According to the symmetry analysis, the allowed NRSF components are $f_{xxx}$, $f_{xyy}$, and $f_{xzz}$.
We plot the temperature dependence of each component in Fig.~\ref{App_Fig_UTe2_NRSF_50_all}.
The NRSF $f_{xyy}$ and $f_{xzz}$ are almost determined by the Fermi-surface effect as we show in the main text for $f_{xxx}$.
By decomposing the Fermi-surface contribution, we identified that the NRSF is almost determined by one of the Fermi-surface terms
		\begin{equation}
                -\frac{1}{2} \sum_a  \vj^\mu_{aa}\vj^\nu_{aa}\vj^\lambda_{aa} \partial^2_\varepsilon f_a.\label{app_dominant_FSurfterm_NRSF}
        \end{equation}
This term plays an essential role in (nearly) nodal superconducting states, although it is strongly suppressed in fully gapped superconductors (see Sec.~\ref{App_Sec_NRSF_gapful_case}). 

\begin{figure}[htbp]
                \centering
                \includegraphics[width=0.50\linewidth,clip]{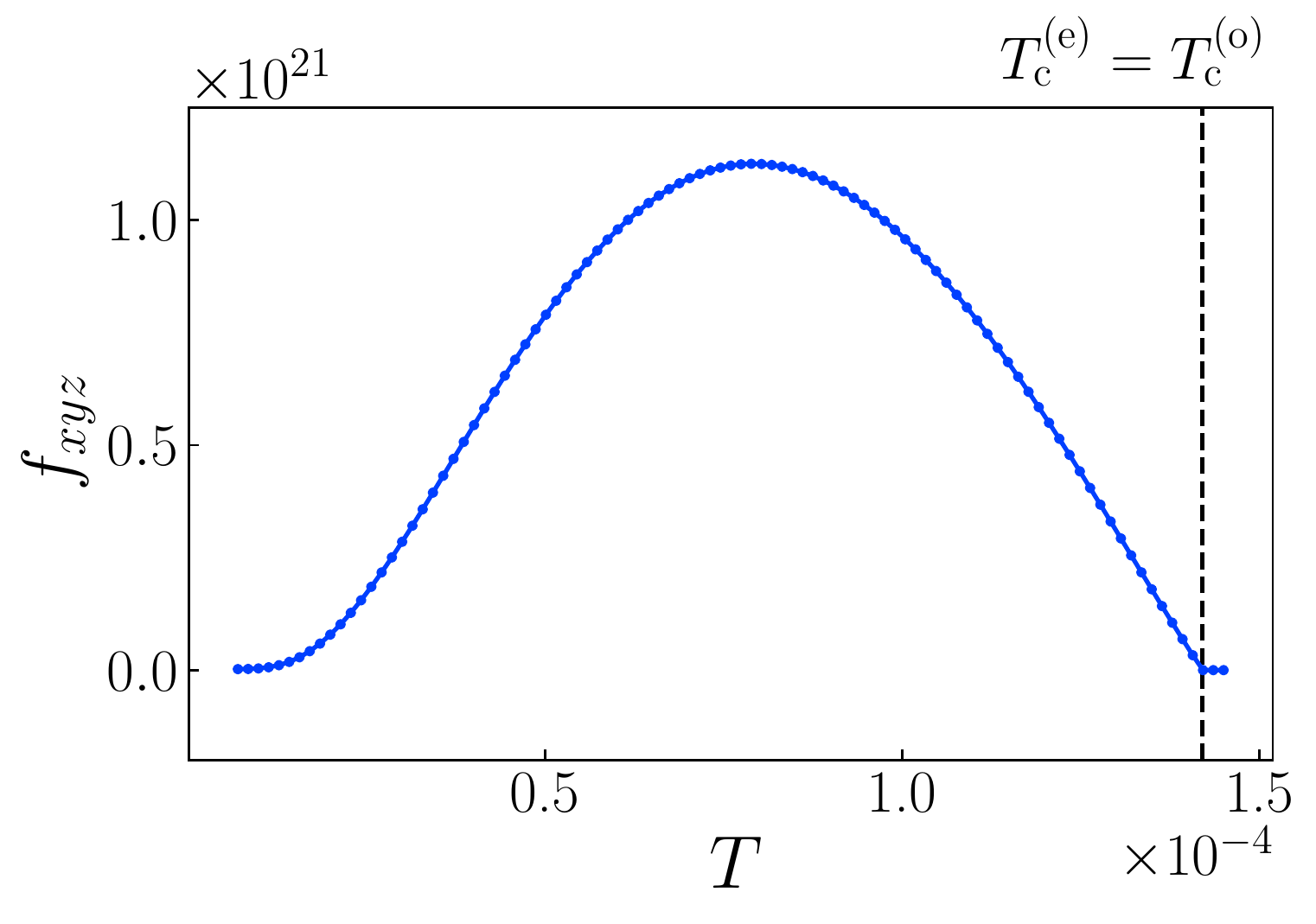}
                \caption{Temperature dependence of the NRSF tensor $f_{xyz}$ in the $A_g + iA_{u}$ equally-pairing ($r=0.5$) state. The unit and parameters for the numerical calculation are the same as Fig.~\ref{App_Fig_UTe2_NRSF_50_all}.}
                \label{App_Fig_NRSF_Au}
                \end{figure}

Following the parallel calculations, we obtain the NRSF for the $A_g+iA_u$ pairing state. The allowed NRSF component is $f_{xyz}$ which indicates that the nonreciprocal Meissner response occurs under the magnetic field perpendicular to the $[111]$-direction. The dependences on the temperature $T$ and the pairing ratio $r$ are found to be qualitatively similar to the $A_g+iB_{3u}$ state. For example, we show the temperature dependence of the NRSF $f_{xyz}$ for the equally-pairing case ($r=0.5$) in Fig.~\ref{App_Fig_NRSF_Au}. The prominent temperature dependence originates from the Fermi-surface term in Eq.~\eqref{app_dominant_FSurfterm_NRSF}.

\section{Nonreciprocal superfluid density with isotropic s-wave pairing}
\label{App_Sec_NRSF_gapful_case}

We studied the model of \utt{} in the main text and Sec.~\ref{App_Sec_UTe2_model}, where we found a sizable Fermi-surface contribution to the NRSF arising from 
the nodal-like gap structure.
In this section, we present another model study to demonstrate that the origin and temperature dependence of the NRSF drastically change in the presence of the isotropic superconducting gap. We discuss two examples. One is the staggered Rashba Hamiltonian. The other is the same Hamiltonian as in the main text and Sec.~\ref{App_Sec_UTe2_model}, although the even-parity pair potential is assumed to be an isotropic $s$-wave one.

\subsection{Staggered Rashba model}
\label{App_Sec_NRSF_staggered_rashba}

The model represents a two-dimensional locally-noncentrosymmetric system. We consider the square lattice and the Rashba-type sublattice-dependent spin-orbit coupling~\cite{Zelezny2014}. The Hamiltonian for the normal state is given by
        \begin{equation}
        H_{\bk}^\text{N} = \left( \varepsilon_0- \mu  \right) + V \rho_x + \bm{g}\cdot \bm{\sigma} \rho_z,\label{App_staggeredR_normal_hamiltonian}
        \end{equation}
where we introduced the Pauli matrices $\bm{\sigma}$ and $\bm{\rho}$ for the spin and sublattice degrees of freedom. Each component reads
        \begin{align}
        &\varepsilon_0 = -4 t_1 \cos{k_x} \cos{k_y},\\
        &V = -2 t_2 \left( \cos{k_x}+  \cos{k_y} \right),\\
        &\bm{g} = 4\alpha \left( \cos{k_x}\sin{k_y},-\sin{k_x}\cos{k_y},0  \right).
        \end{align}
The model parameters are chosen as
        \begin{equation}
        \mu=-4,~\alpha= 0.3,~t_1 = 0.6,~t_2 = 1.0.\label{staggeredR_model_parameters}
         \end{equation}
The mixed-parity superconducting pair potential is introduced by the molecular field
        \begin{equation}
        \hat{\Delta}_\bk =  \left(  \psi_\bk + i \bm{d}_\bk\cdot \bm{\sigma}\right) i\sigma_y \rho_0. 
        \end{equation}
The spin-singlet ($\psi_\bk$) and spin-triplet ($\bm{d}_\bk$) components respectively correspond to the even and odd-parity pairings, and we assume
        \begin{equation}
        \psi_\bk = \Delta_\text{e},~\bm{d}_\bk =\Delta_\text{o} \sin{k_x}\cos{k_y} \hat{z}.
        \end{equation}
The symmetry of the parity-mixed superconductivity is denoted by the $A_g + i E_u$ representation of a tetragonal point group $D_{4h}$. Accordingly, the allowed NRSF components are given as in the case of the $A_g+ i B_{3u}$ pairing of \utt{}. In the following numerical calculations, we set the total amplitude of the pair potentials to $\Delta_\text{e+o} = |\Delta_\text{e}|+|\Delta_\text{o}| = 0.1$. We adopt a series of the ratio $r$ defined by $|\Delta_\text{e}|=r\,\Delta_\text{e+o}$.

\begin{figure}[htbp]
                \centering
                \includegraphics[width=0.9\linewidth,clip]{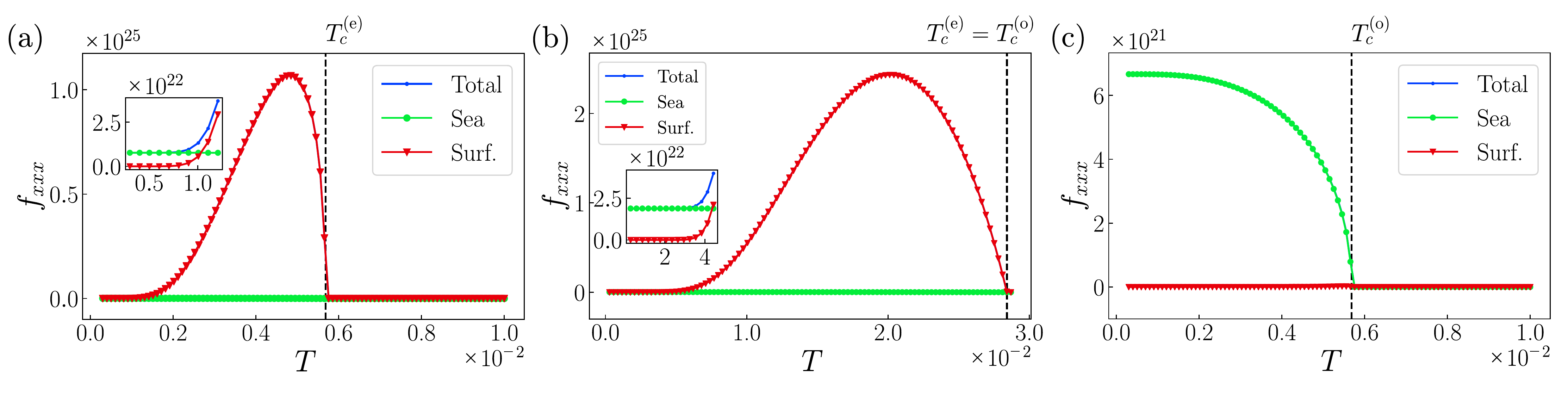}
                \caption{Temperature dependence of the NRSF $f_{xxx}$ (A$\cdot$V$^{-2}\cdot$s$^{-2}$) in the staggered Rashba model. The total NRSF (blue) is decomposed into the Fermi-sea (green) and Fermi-surface (red) contributions. (a) $p$-wave (odd-parity) dominant regime ($r=0.1$). (b) Equally-pairing regime ($r=0.5$). (c) $s$-wave (even-parity) dominant regime ($r=0.9$). The horizontal axis of the insets is $T \times 10^3$. The dashed lines guide the transition temperature of even-parity ($T_\text{c}^\text{(e)}$) and odd-parity ($T_\text{c}^\text{(o)}$) superconductivity. We use the $N^2=1500^2$ discretized Brillouin zone in the numerical calculation.}
                \label{App_Fig_staggeredR_NRSF_SCratio_Tdep}
                \end{figure}

The model illustrates the crossover from a gapful superconductor ($r=1$) to a nodal superconductor ($r=0$). Thus, we can see how quasiparticle excitation contributes to the NRSF by changing the ratio $r$.
The temperature dependence of the pair potentials is phenomenologically assumed by Eq.~\eqref{App_phenomenological_SC_temperature_dependence}. Figure~\ref{App_Fig_staggeredR_NRSF_SCratio_Tdep} shows the NRSF with the ratio $r=0.1,~0.5,~0.9$. For a moderate weight of $p$-wave pairing ($r=0.1,~0.5$), the Fermi-surface contribution plays a major role. On the other hand, in the $s$-wave dominant regime ($r=0.9$), the Fermi-sea contribution is dominant, while the Fermi-surface contribution is negligible due to the absence of thermal excitations. Such behavior is also found in the low-temperature regime for $r=0.1,~0.5$ [see insets of Figs.~\ref{App_Fig_staggeredR_NRSF_SCratio_Tdep}(a) and \ref{App_Fig_staggeredR_NRSF_SCratio_Tdep}(b)]. The NRSF is remarkably enhanced by the Fermi-surface contribution, which can be more than $10^3$ times larger than the Fermi-sea contribution when thermal excitations moderately occur.
Therefore, the NRSF in anisotropic superconductors can be much larger than that in the isotropic superconductors.

\subsection{Uranium ditelluride model with isotropic $s$-wave pairing}
\label{App_Sec_gapful_UTe2}

Instead of the extended $s$-wave pairing $\psi_\bk=\Delta_\text{e} \cos{k_x}$ adopted in the main text and Sec.~\ref{App_Sec_UTe2_model}, we take the isotropic form $\psi'_\bk=\Delta_\text{e}$. Except for the $s$-wave pair potential, we adopt the same Hamiltonian as in Sec.~\ref{App_Sec_UTe2_model}. 

\begin{figure}[htbp]
                \centering
                \includegraphics[width=0.9\linewidth,clip]{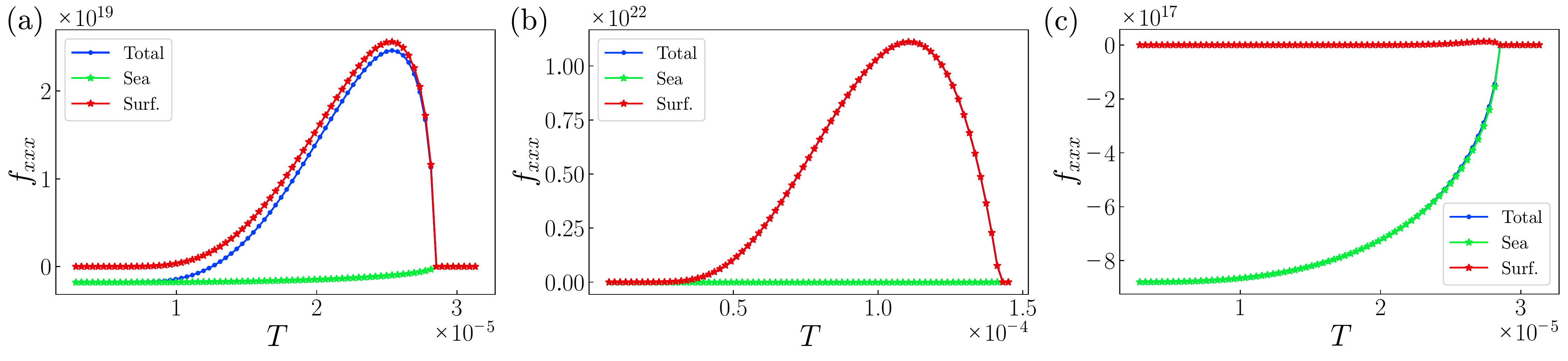}
                \caption{Temperature dependence of the NRSF $f_{xxx}$ in the \utt{} model with the isotropic $s$-wave pairing. The convention is the same as in Fig.~\ref{App_Fig_staggeredR_NRSF_SCratio_Tdep}. (a) $p$-wave (odd-parity) dominant regime ($r=0.1$). (b) Equally-pairing regime ($r=0.5$). (c) $s$-wave (even-parity) dominant regime ($r=0.9$). We use the $N^3=150^3$ discretized Brillouin zone in the numerical calculation.}
                \label{App_Fig_gapful_ute2_NRSF_SCratio_Tdep}
                \end{figure}

Figure~\ref{App_Fig_gapful_ute2_NRSF_SCratio_Tdep} shows the temperature dependence of the NRSF $f_{xxx}$ when the several pairing ratios $r=0.1,0.5,0.9$ are taken. Similarly to Fig.~\ref{App_Fig_staggeredR_NRSF_SCratio_Tdep}, the NRSF is strongly enhanced by the Fermi-surface term and reaches the maximum value in the intermediate temperature regime for the the cases with $r=0.1$ and $0.5$ [Figs.~\ref{App_Fig_gapful_ute2_NRSF_SCratio_Tdep}(a) and \ref{App_Fig_gapful_ute2_NRSF_SCratio_Tdep}(b)]. On the other hand, the Fermi-surface term plays a minor role in the $s$-wave dominant state with $r=0.9$ [Fig.~\ref{App_Fig_gapful_ute2_NRSF_SCratio_Tdep}(c)], because the nearly isotropic superconducting gap suppresses the quasiparticle excitation. Instead, the Fermi sea determines the NRSF but makes a much smaller contribution than the Fermi-surface effect does in the presence of a sizable $p$-wave component.

In conclusion, the structure of the superconducting gap significantly influences the nonreciprocal property of the superfluid density due to a sizable contribution from the quasiparticle excitations. Thus, the temperature dependence of the NRSF will be informative in determining the superconducting gap structure. For instance, if we carefully control the temperature $T$ and pairing ratio $r$, the NRSF and the associated nonreciprocal Meissner response show a drastic change.

\newpage

\bibliography{paper_N_Meissner}

\end{document}